\documentclass[preprint,floatfix,prl,preprintnumbers,footinbib]{revtex4}
\usepackage{amssymb}
\usepackage[dvips]{graphicx}
\usepackage{dcolumn}
\usepackage{bm}
\usepackage[usenames]{color}

\input{tcilatex}

\begin{document}

\title{Co-existence of the Meissner and vortex-state on a superconducting
spherical shell}
\author{J. Tempere$^{1,2}$, V. N. Gladilin$^{1,3}$, I. F. Silvera$^{2}$, J.
T. Devreese$^{1}$, and V. V. Moshchalkov$^{3}$}
\affiliation{$^{1}$ TFVS, Universiteit Antwerpen, Groenenborgerlaan 171, 2020 Antwerpen,
Belgium }
\affiliation{$^{2}$ Lyman Laboratory of Physics, Harvard University, Cambridge,
Massachusetts 02138, USA }
\affiliation{$^{3}$ INPAC, K. U. Leuven, Celestijnenlaan 200 D, B-3001 Leuven, Belgium }

\begin{abstract}
We show that on superconducting spherical nanoshells, the co-existence of
the Meissner state with a variety of vortex patterns drives the phase
transition to higher magnetic fields. The spherical geometry leads to a
Magnus-Lorentz force pushing the nucleating vortices and antivortices
towards the poles, overcoming local pinning centers, preventing
vortex-antivortex recombination and leading to the appearance of a Meissner
belt around the sphere's equator. In sufficiently small and thin spherical
shells paramagnetic vortex states can be stable, enabling spatial separation
of freely moving shells with different radii and vorticity in an
inhomogeneous external magnetic field.
\end{abstract}

\date{\today}
\maketitle

Controlling and understanding vortex behavior and creating a guided vortex
(fluxon) motion in superconductors is crucial for developing new fluxonics
devices~\cite{1,2,3,4,hastings2003,zhu04,5}. Several key experiments have
demonstrated how different vortex patterns can be created and guided in
mesoscopic and nanoscopic superconductors~\cite{1,2,3,4}. These
breakthroughs in the pursuit of 'fluxonics' have focused on hybrid
superconductor/ferromagnet nanosystems~\cite{1,2}, or on the use of
nanostructured superconductors~\cite{3,4, field02}. Here, we investigate how
the geometry (curvature and topology) of the superconducting layer rather
than the patterning can be used to control flux. Curvature and surface
topology, which are known to strongly affect, e.g., charge ordering and the
dynamics of defects on spherical surfaces~\cite{bowick1, bowick2}, have a
profound influence also on the vortex behavior in superconducting layers.
Curvature also affects the superfluid transition, notably in helium films
adsorbed in porous materials with small grain sizes\cite{WilliamsPRB33},
where the vortex dynamics can be driven experimentally by rotation\cite%
{FukudaJLTP113}. In this paper we focus on the vortex behavior in spherical
superconducting nanoshells. These are nanoparticles consisting of a
dielectric core of typically 50-200 nm in radius, coated by a 5-20 nm thin
metallic shell~\cite{6}. Thermodynamically stable vortex states in
nanoshells as well as the dynamics of vortex trapping and releasing are
investigated within the Ginzburg-Landau formalism applied to a spherical
surface. Results are presented of both variational analysis and numerical
solutions.

In the Ginzburg-Landau formalism, superconductors are described by a
macroscopic wave function $\psi=|\psi|\exp(i\varphi)$ that couples to the
electromagnetic field and takes the role of the complex order parameter for
the superconducting phase. The modulus square of the order parameter $%
|\psi|^{2}$ corresponds to the density of Cooper pairs, whereas the gradient
of its phase $\varphi$ defines the supercurrent. We focus on thin
superconducting nanoshells with a shell thickness $\mathcal{W}$ smaller than
the correlation length $\xi$ (that also defines the vortex core size). This
requirement simplifies the treatment in two important ways. First, the order
parameter will be constant in the shell along the radial direction, so $\psi$
will only depend on the spherical angles $\Omega=\{ \theta,\phi \}$. Second,
if the thickness of a shell also satisfies the inequality $\mathcal{WR}\ll
\lambda^{2}$, where $\mathcal{R}$ is the shell radius and $\lambda$ is the
London penetration depth, the external magnetic field will be only weakly
perturbed by the nanoshell.

\begin{figure}[tbp]
\centering \includegraphics[width=8.5cm]{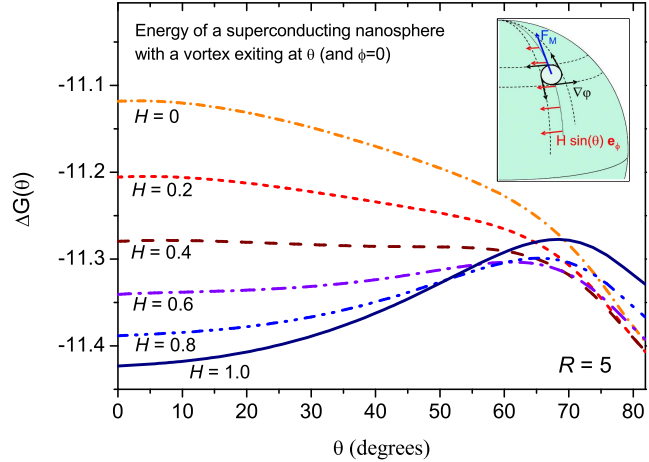}
\caption{(color online). The Gibbs free energy of a vortex-antivortex pair,
displaced away from the poles down to a latitude $\protect\theta$, is shown
for different values of the applied magnetic field. The origin of the
metastability barrier, which develops with increasing the applied magnetic
field, is a Magnus-Lorentz $F_{M}$ force pushing vortices towards the poles,
as illustrated in the inset. This force arises from a differential velocity
field across the vortex due to the interplay between the supercurrent of the
vortex, proportional to the phase gradient $\protect\nabla \protect\varphi$,
and the screening supercurrent induced by the magnetic field $H$. }
\label{fig1}
\end{figure}

Bulk superconductors expel the magnetic field and form a Meissner state up
to a lower critical field $H_{c1}$. When exposed to higher fields, type II
superconductors allow the magnetic field to penetrate in the form of
quantized superconducting Abrikosov vortices, up to a field $H_{c2}$. The
behavior of superconducting nanoshells can be derived from a variational
argument in which we consider a nanoshell with a vortex line along the $z$%
-axis at a magnetic field $H_{c1}<H<H_{c2}$. This vortex line punctures the
shell in two points, forming `cores' around which the two-dimensional
superflow on the surface takes place. The two-dimensional (2D) superflow on
the northern hemisphere ($\theta<\pi/2$) rotates anticlockwise around the
unit vector $\mathbf{e}_{r}$ at the core, whereas on the southern hemisphere
($\theta>\pi/2$) it rotates clockwise around the unit vector $\mathbf{e}_{r}$
at the southern core. We will refer to the local flow pattern on the
northern hemisphere as a 2D-vortex and to that on the southern hemisphere as
a 2D-antivortex. When we ramp down the magnetic field to a value $H<H_{c1}$,
where the vortex state is unstable, the vortex line, still parallel to the
south-north axis, is moved away from the poles (so as to expel vorticity
from the system). In other words, the 2D vortex and 2D antivortex move on
their respective hemispheres towards the equator. There, they can merge, and
the clockwise and anticlockwise flows cancel each other out, leaving a
uniform order parameter. The dynamical behaviour for expelling vorticity,
sketched above in a qualitative way, can be investigated more rigorously
using variational calculus on the Gibbs free energy. In the case of thin
shells, $\mathcal{W}\ll \lambda^{2}/\mathcal{R}$, the Gibbs free energy
becomes 
\begin{eqnarray}
\Delta G & =\int d\Omega \text{ }\left \{ \left( \mathbf{\nabla}_{\Omega
}\left \vert \psi \right \vert \right) ^{2}+\left \vert \psi \right \vert
^{2}\left[ \mathbf{\nabla}_{\Omega}\varphi-H\sin(\theta)\mathbf{e}_{\phi }%
\right] ^{2}\right.  \nonumber \\
& \left. -2{R}^{2}\left \vert \psi \right \vert ^{2}\left( 1-\frac{1}{2}%
\left \vert \psi \right \vert ^{2}\right) \right \}   \label{gibbs}
\end{eqnarray}
In this expression, we use spherical coordinates with the $z$-axis parallel
to the external magnetic field such that $\mathbf{\nabla}_{\Omega}=\mathbf{e}%
_{\theta}(\partial/\partial \theta)+\mathbf{e}_{\phi}\sin^{-1}(\theta
)(\partial/\partial \phi)$. Two experimentally tunable parameters remain:
the radius of the nanoshell, and the external magnetic field. The external
magnetic field $\mathcal{H}$ appears in (\ref{gibbs}) as ${H}=\Phi/\Phi
_{0}=\pi \mathcal{R}^{2}\mathcal{H}\Phi_{0}$ corresponding to the amount of
flux quanta of the applied field that pass through the equatorial plane of
the sphere. The radius of the shell appears as ${R}=\mathcal{R}/(\sqrt{2}\xi)
$, the ratio of shell radius to the coherence length multiplied by $\sqrt{2}$%
. To describe a vortex state, where the core of the vortex on the northern
hemisphere is at $\Omega=\{ \theta_{v},0\}$ and the corresponding antivortex
is at $\{ \pi-\theta_{v},0\}$ on the southern hemisphere, we use the trial
wave function 
\begin{eqnarray}
\left \vert \psi(\theta,\phi)\right \vert & =a\left( 1-e^{-{R}\vartheta
}\right) \left( 1-e^{-{R(}\pi-\vartheta)}\right)  \nonumber \\
\varphi(\theta,\phi) & =\arctan \left( \frac{\sin \theta \sin \phi}{\sin
\theta \cos \phi-\sin \theta_{v}}\right) ,  \nonumber \\
& -\arctan \left( \frac{\sin \theta \sin \phi}{\sin \theta \cos \phi-\sin
(\pi-\theta_{v})}\right) ,  \nonumber
\end{eqnarray}
where $a$ is a variational parameter and 
\[
\cos \vartheta=\cos \theta \cos \theta_{v}+\sin \theta \sin
\theta_{v}\cos(\phi). 
\]
As illustrated in Fig.~\ref{fig1}, where the calculated Gibbs free energy is
plotted for different values of the magnetic field and for the particular
case of ${R}=5$, the homogeneous external magnetic field gives rise to an
energy barrier that pushes the 2D vortex and the 2D antivortex away from the
equator and towards the poles, separating the pair. This is in remarkable
contrast with flat superconducting films in a homogeneous magnetic field,
where vortex-antivortex pairs tend to annihilate. Unlike the Bean-Livingston
barrier\cite{beanlivingston}, the present metastability barrier is not
caused by the interaction with an image vortex but by the surface curvature.
The inset of Fig.~\ref{fig1} illustrates the origin of the Magnus-Lorentz
force, responsible for separation of vortex-antivortex pairs. When multiple
single vortices and antivortices are present, they aggregate at the opposite
poles, forming a vortex lattice polar region.

\begin{figure}[tbp]
\centering \includegraphics[width=8.5cm]{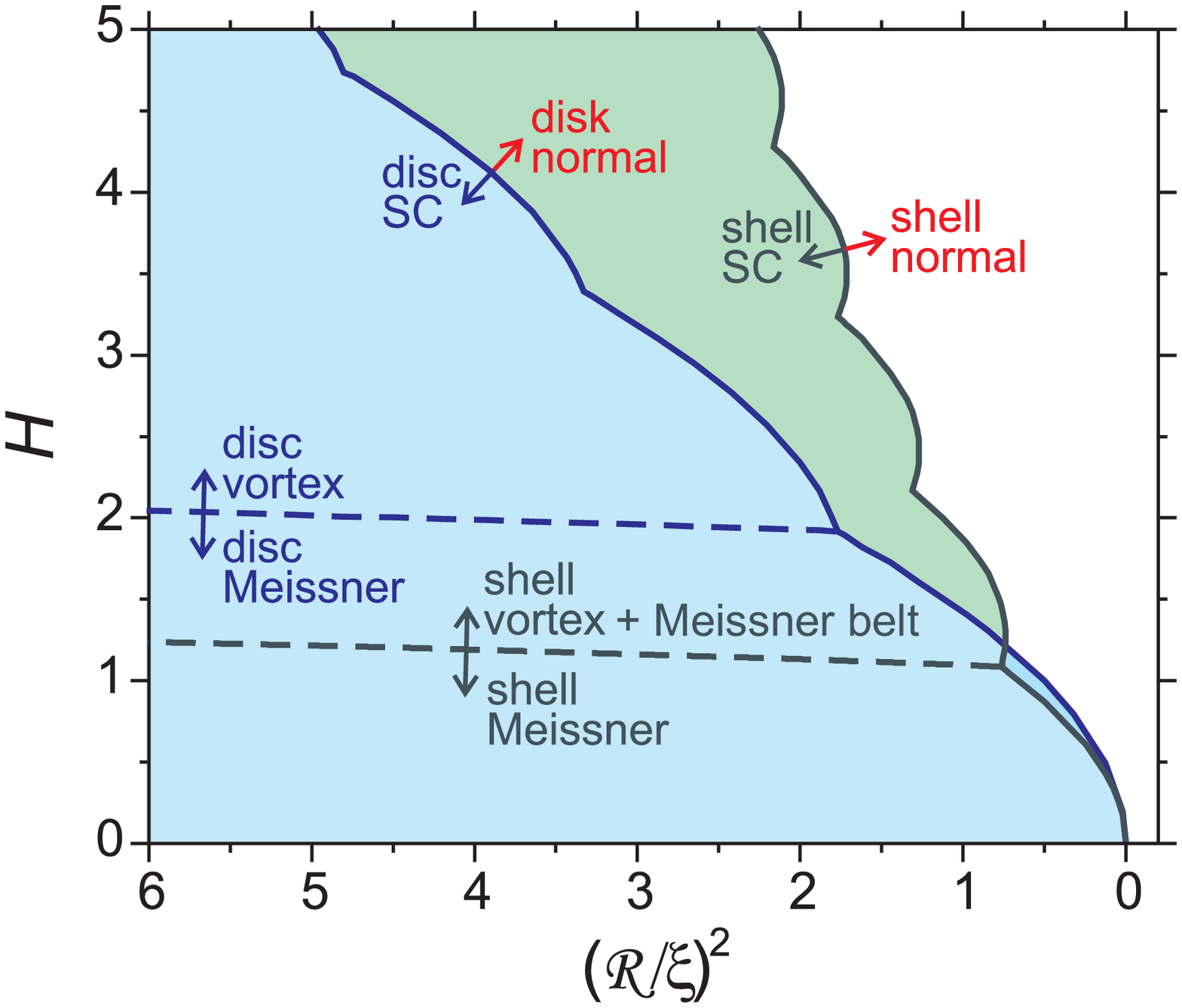}
\caption{(color online). The phase diagram for a thin spherical shell is
compared to that of a flat thin disk. The shaded regions show the values of
magnetic field and radius where the superconducting (SC) state is supported.
The solid lines correspond to the boundaries between the thermodynamically
stable normal and superconducting states. The dashed lines correspond to the
boundaries between the thermodynamically stable Meissner and vortex states.}
\label{fig2}
\end{figure}

The equatorial region remains a vortex-free 'Meissner state', although a
shielding current is present. The Meissner belt at the equator represents a
Cooper pair reservoir tangent to the magnetic field. As shown in Ref.~\cite%
{8a}, the current-carrying capacity of superconducting strips can be
enhanced by geometrical barriers, which result in the co-existence of
isolated vortex-filled regions with current-carrying vortex-free Meissner
regions. In the case of a nanoshell, we find that the co-existence of the
Meissner belt with the vortex lattice at the poles aids superconductivity in
a similar way. In Fig.~\ref{fig2} we compare the superconductivity phase
diagram for nanoshells and disks. The results, shown in this figure and
below, originate from a finite-element numerical solution of the
time-dependent Ginzburg-Landau (TDGL) equation, described in~\cite{prb2008}
and applicable also to the case when magnetic fields, induced by
supercurrents, are non-negligible. The set of parameters, governing the
solution of the TDGL equation in~\cite{prb2008}, contains -- in addition to $%
R$ and $H$ -- also the ratio $\mathcal{WR}/\lambda^{2}$. Figure ~\ref{fig2}
corresponds to the case of thin layers where $\mathcal{WR}\ll \lambda^{2}$.
As seen from Fig.~\ref{fig2}, for nanoshells the region in the phase diagram
where superconducting vortex state is present is considerably expanded:
superconducting nanoshells tolerate considerably higher magnetic fields than
superconducting disks.

\begin{figure}[tbp]
\centering \includegraphics[width=8.5cm]{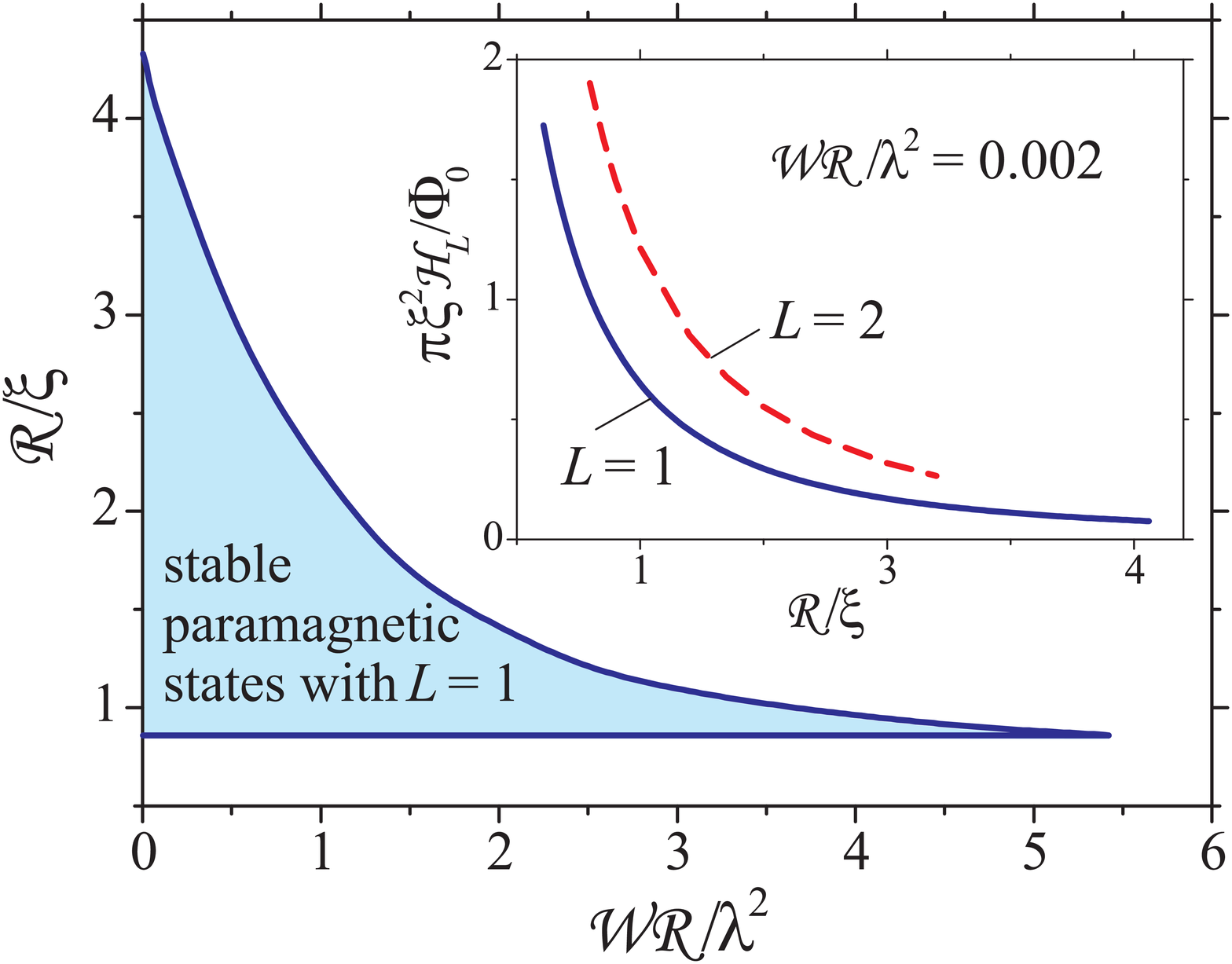}
\caption{(color online). Boundaries of the region, where paramagnetic states
with $L=1$ on a spherical shell of radius $\mathcal{R}$ and thickness $%
\mathcal{W}$ can be thermodynamically stable. Inset: Magnetic fields, which
correspond to the equilibrium positions of thin spherical shells with
vorticity $L=1$ and $L=2$ in an inhomogeneous magnetic field, as a function
of the shell radius. }
\label{fig5}
\end{figure}

Due to the presence of the Meissner belt, in relatively large shells not
only the Meissner state but also the thermodynamically stable states with
one or few quanta of vorticity are characterized by a negative magnetic
moment of the shell, i.e. these states are always diamagnetic (see e.g.
Fig.~2 in Ref.~\cite{prb2008}). Our calculations show that paramagnetic
states with one or few vortex pairs (or a pair of giant vortices) can be
thermodynamically stable only in sufficiently small and thin shells. In Fig.~%
\ref{fig5}, this is illustrated for the case $L=1$, where $L$ is the number
of quanta of vorticity present. For small and thin spherical shells, the
magnetic fields $\mathcal{H}_{L}$, which correspond to the minima of the
free energy for states with $L=1,2,\ldots$, lie in the range where these
states are thermodynamically stable. As a result, those shells -- when
assumed to be able to move freely -- can manifest a rather peculiar behavior
in a weakly inhomogeneous magnetic field (i.e. in a field, which
substantially varies only on a size scale much larger than the shell
radius). Indeed, as illustrated in the inset to Fig.~\ref{fig5}, the values $%
\mathcal{H}_{L}$ (which are almost insensitive to the shell thickness $%
\mathcal{W}$) strongly depend on $\mathcal{R}$. This means that shells with
the same nonzero vorticity but different radius (as well as shells with the
same size but different vorticity) can be spatially separated in an
inhomogeneous magnetic field. In a sense, this situation is analogous to the
quantized levitation, analyzed in Ref.~\cite{fink1996} for a superconducting
ring in the magnetic field of another, fixed ring (of course, while for the
levitating ring one should take care of keeping its orientation parallel to
the fixed ring, there is no need of such a care in the case of spherically
symmetric shells). The aforedescribed behavior of thin spherical nanoshells
is in a remarkable contrast to the case of full spherical grains. As implied
by the results~\cite{Baelus}, based on the linearized Ginzburg-Landau
equation, as well as by our calculations for the non-linear TDGL equation,
thermodynamically stable states in full spherical grains are always
diamagnetic. This means that in an inhomogeneous magnetic field the
thermodynamic equilibrium position of all full grains will correspond to $%
\mathcal{H}=0$ (or to the lowest available value of $\mathcal{H}$).

Vortices can only be expelled, or nucleated, near the equator, where the
metastability energy barrier is high, enabling flux hysteresis. In Fig.~\ref%
{fig3}, the black solid curves show the Gibbs free energy of the ground
state as a function of the magnetic field, for a given shell geometry. The
Meissner state is thermodynamically stable below $H\approx1.5$, while for $%
1.5<H<2.85$, the state with a single vortex is stable. When the magnetic
field is slowly ramped up (green dashed curve), and down again (red dotted
curve) a clear hysteresis effect in the vorticity is seen. Due to the
metastability barrier, when the field is ramped up, the vortex line is
prohibited from entering the nanoshell and vortices start nucleating at the
equator only at $H\approx4.15$ when the shell makes a transition from the
Meissner state to the state with three circulation quanta. To expel all
vortices from the shell, an external magnetic field has to be lowered down
to $H\approx0.5$. Our calculations show that flux hysteresis is enhanced
when increasing the thickness and size of the shell: at $\mathcal{WR}\sim
\lambda^{2}$ an external magnetic field of opposite direction should be
applied in order to remove flux completely from the shell. We emphasize that
here the flux is trapped not by flux pinning at imperfections, but rather by
the topology of the system itself. While the above simulations are performed
for idealized spherically symmetric nanoshells, in realistic nanoshells,
inevitable imperfections may perturb the trapping potential for vortices. In
order to model the effect of those imperfections, we have considered
nanoshells with spatial variations of the Ginzburg-Landau parameter $\kappa$%
. According to the results of our calculations, though inhomogeneity of $%
\kappa$ usually tends to destabilize metastable vortex states, this
destabilizing effect on vortex trapping is not dramatic in the case of
relatively small variations of $\kappa$. Our results imply that vortex
trapping should be robust also with respect to moderate deviations of the
nanoshell shape from sphericity.

\begin{figure}[tbp]
\centering \includegraphics[width=8.5cm]{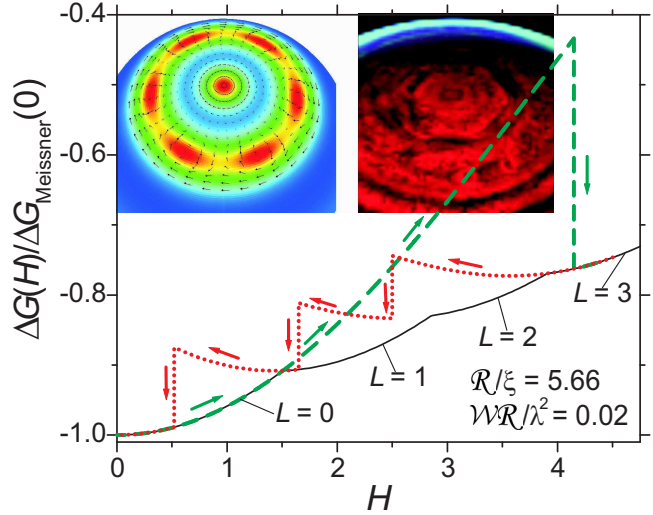}
\caption{(color online). Flux hysteresis in a superconducting nanoshell. The
Gibbs free energy difference between the normal and superconducting state is
plotted as a function of the magnetic field. The solid curve shows the
thermodynamically stable state. The dashed (dotted) curve corresponds to a
slow increase (decrease) of the applied magnetic field. Inset: On the left,
a vortex pattern that arises at high magnetic field ($H=20$) when a
ring-like vortex in a nanoshell with $\mathcal{R}/\protect\xi=11.5)$ and $%
\mathcal{WR}\ll \protect\lambda^{2}$ breaks up into separate vortices. The
color scale indicates the Cooper pair density, from blue (high) to red
(low). The supercurrents are indicated by the arrow field. On the right, a
cloud pattern observed at the north pole of Saturn (Credit:
NASA/JPL/University of Arizona).}
\label{fig3}
\end{figure}

The particular dynamics of vortices entering the shell is shown in Fig.\ref%
{fig4}. The distribution of supercurrents, typical for a pair of 2D
vortices, appears only when the separation between the vortex cores is of
the order of the twice the coherence length. As a consequence, vortex cores
can only be present outside a Meissner belt of latitudes $\pm \Delta \theta
_{\mathrm{M}}\approx \mathrm{arctan}[\xi/(2\mathcal{R})]$ around the
equator. Qualitatively, this Meissner belt resembles the hurricane-free belt
of 3$^{\circ}$ latitude around the Earth equator. Expressing the
Ginzburg-Landau equations in hydrodynamic form, the resulting equations for
superfluid velocity and Cooper pair density are formally similar to the
shallow-atmosphere Euler equations used to model atmospheric dynamics~\cite%
{7}. As a result, there is a similarity between the behavior of atmospheric
vortices (cyclones) on the macroscopic globe and superconducting vortices on
the nanoshell. This is illustrated in the inset to Fig.~\ref{fig3} for the
formation of polar vortex lattices. The lhs panel of this inset shows a
vortex pattern that arises on a nanoshell when a ring-like vortex breaks up
into separate vortices. The rhs panel of the inset depicts a cloud pattern
observed at the north pole of Saturn. The initial axially symmetric state in
the nanoshell included nonuniform vorticity~\cite{prb2008,Zhao} with a
different angular momentum state near the poles and near the equator.
Differential wind speeds (or superconducting currents) in two bands circling
the pole lead to a depression (of pressure in the atmosphere, and of Cooper
pair density in the superconductor) in the interface between the bands. A
modulation of this depression reduces the energy in the case of a
superconductor - one can speculate that a similar mechanism may be at work
at Saturn's pole.

\begin{figure}[tbp]
\centering \includegraphics[width=8.5cm]{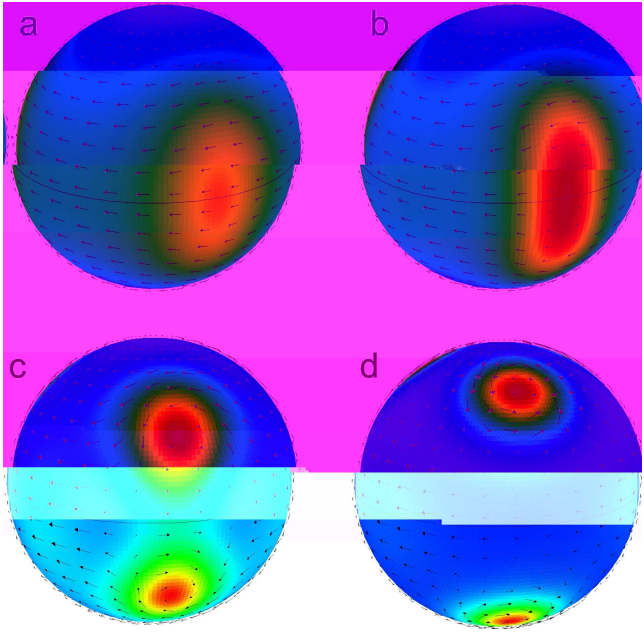}
\caption{(color online). Four snapshots of the dynamical process of vortices
entering a superconducting nanoshell with $\mathcal{R}/\protect\xi=5.66$ and 
$\mathcal{WR}/\protect\lambda^{2}=0.02$ at $H=4.15$. The color scale
indicates the Cooper pair density, from blue (high) to red (low). The
supercurrents are indicated by the arrow field. A redistribution of the
Cooper pair density is already present in the Meissner state (a). The
vortices nucleate at a depression of the Cooper pair density at the equator
(b), separate into two counterrotating two-dimensional vortices (c) and
proceed towards the poles (d).}
\label{fig4}
\end{figure}

In nanoscopic superconductors, confinement potentials and periodic
modulation of material parameters have been explored as tools to manipulate
flux and quantum coherence. Here, we have shown that also the geometry of
the sample can be used to manipulate flux. The spherical shell geometry
leads to two important properties: first, 2D-vortex-antivortex pairs tend to
separate rather than annihilate, and second, the curvature enables the
co-existence of a vortex state and a Meissner (non-vortex) belt close to the
equator on the same surface. These properties result in a higher critical
magnetic field in a shell in comparison to a disc with corresponding
cross-section. Also we find a pronounced hysteresis effect for flux trapping
in the nanoshell, allowing magnetic separation of spheres with different
vorticity in an inhomogeneous field. Experimental techniques for producing
monodisperse and uniform SiO$_{2}$ nanospheres \cite{stob} that can be
coated with a metal \cite{6,nanosphere} such as niobium, either
individually, or in a film of hemispheres, offer the prospect to probe the
enhanced magnetic properties of nanoshells discussed in the present work.

\begin{acknowledgments}
The authors would like to thank G. Williams for pointing out the analogy
between vortex behavior in superconducting nanoshells and vortex behavior in
liquid helium films in porous materials\cite{WilliamsPRB33,FukudaJLTP113}.
This work was supported by the Fund for Scientific Research-Flanders
Projects G.0370.09N, G.0180.09N, G.0356.06, G.0115.06, G.0435.03, the WOG
Project WO.033.09N, and the U.S. Department of Energy, Grant No.
DE-FG02-ER45978. V.V.M. acknowledges Methusalem Funding by the Flemish
Government. J.T acknowledges financial support from the Special Research
Fund of the University of Antwerp, BOF NOI UA 2004.
\end{acknowledgments}

\end{document}